\documentclass[12pt]{article}
\usepackage{amsfonts}
\usepackage{latexsym}
\usepackage{psfig}
\title{ A Model of Interface Growth \linebreak 
with non-Burgers Dynamical Exponent}
\author{ Hari M. Koduvely$^\dagger$ and Deepak Dhar$^\ddagger$  \\
Theoretical Physics Group \\
Tata Institute of Fundamental Research \\
Mumbai 400 005, India.}
\begin{document}
\maketitle \abstract{We define a new model of interface roughening
which has the property that the minimum of interface height is
conserved locally during the growth. This model corresponds to the
limit $q \rightarrow \infty$ of the $q$-color dimer
deposition-evaporation model introduced by us earlier [Hari Menon M K
and Dhar D 1995 {\it J. Phys. A: Math. Gen.} {\bf 28} 6517].  We
present numerical evidence from Monte Carlo simulations and the exact
diagonalization of the evolution operator on finite rings that this
model is not in the universality class of the Kardar-Parisi-Zhang
interface growth model. The dynamical exponent $z$ in one dimension is
larger than $2$, with $z\approx 2.5$. And there are logarithmic
corrections to the scaling of the gap with system size. Higher
dimensional generalization of the model is briefly discussed.}
\newpage
\section{Introduction}
The structure of growing interfaces has been a major subject of study
in recent years \cite{stbb,hhz}. Most of these studies concern the
large scale properties of a moving interface between two phases. The
recent spurt of interest in this subject started with the proposal of
a nonlinear evolution equation for the interface height $h(\vec x,t)$
by Kardar, Parisi and Zhang (KPZ) one decade ago \cite{kpz}. KPZ
argued that consistent with the symmetries of the interface, the
lowest order nonlinear term in the growth equation is proportional to
$(\nabla h)^2$. Hence in a large number of physical situations the
growing interface can be described by a noisy Burgers equation. In
$1+1$ dimensions this nonlinear term is the most relevant
perturbation, and it gives rise to a dynamical universality class
characterized by the value of the dynamical exponent $z=3/2$. Addition
of higher order nonlinear terms, consistent with the symmetries, will
not change the value of $z$.

Though this is indeed the case found in many physical situations,
there are cases where the correlations of the interface are not
described by KPZ exponents \cite{snep, krug}. For example, if there
are some constraints present for the growth, we can expect the
relaxation of the interface to become slower.
 
Some generalizations of the KPZ equation where the scalar height variable
is replaced by an $N$-component vector \cite{doh}, or an $N\times N$
matrix \cite{kard} have been studied. But in both these cases it was
found that the value of $z$ remains unchanged in $1+1$ dimensions.

In this context, it is interesting to note that recently a class of
deposition-evaporation models has been studied which in $1+1$
dimensions can be mapped to interface roughening models where the
height variables are $2\times 2$ matrices [8-14]. Monte Carlo
simulations and numerical diagonalization studies have shown that
these models do not belong to the KPZ dynamical universality class
\cite{ddhmk1,ddhmk2}. The numerical studies strongly suggests that
\hbox{$z\approx5/2$}. An example of this is the trimer
deposition-evaporation (TDE) model \cite{mbg1}. Numerical
diagonalization studies on small lattices shows that $z\approx 2.5$
for this model \cite{ddhmk1}. Another example is the q-color dimer
deposition-evaporation (qDDE) model \cite{ddhmk2}. For the case of
$q=2$, qDDE model reduces to the Heisenberg model which is exactly
solvable \cite{mbg1}, and the corresponding value of $z$ is $2$. For
$q > 2$, numerical studies suggests that this model is in the same
universality class as the TDE model.

Though there is fairly strong numerical evidence for this new
universality class with \hbox{$z\approx 2.5$}, so far there have been
no models where the precise value of the dynamical exponent can be
established more convincingly. Obukhov et al have given scaling
arguments that \hbox{$z=5/2$} in the context of a related model for
diffusion of ring polymers in gels \cite{obuk}. Using similar, but
somewhat more careful arguments, Alon and Mukamel reached the same
conclusion \cite{uri}. It is thus of some interest to look for simpler
models in this universality class, for which analytical results can be
obtained. It is well known that many spin models become simpler, and
in some cases even exactly solvable, in the limit when the number of
components of the spin tends to infinity \cite{brez}. This has
motivated us to the study of the qDDE model in the limit of $q
\rightarrow \infty$. Our main result is that in this limit, the qDDE
model reduces to a simpler interface growth model with scalar
heights. {\it This interface model has an interesting constraint that
the minimum height of the interface is conserved locally during the
growth.} We have studied this simpler model numerically, but have not
been able to solve it analytically so far.

This paper is organized as follows: In section $2$ we define the
interface growth model, and write down the stochastic matrix as the
Hamiltonian of a quantum mechanical spin chain. We argue that
conservation of minimum height locally in the model makes the
relaxation slow. In section $3$ we calculate the average height of the
interface in the steady state for a ring of size $L$, and show that
this grows as $L^{1/2}$ for large $L$. In section $4$ we
recapitulate the definition of the qDDE model and briefly list its
known properties. In section $5$ we study the dynamics of the qDDE
model in the limit of large $q$. We show that in this limit the model
simplifies, and for time scales much greater than $1/q$ the model can
be described by an effective Markovian dynamics. In section $6$ we
establish an equivalence between this effective dynamics and the
dynamics of the interface model defined in section $2$.  In section
$7$ we study this interface model using both Monte Carlo simulations
and numerical diagonalization of the stochastic matrix for small
lattice sizes. These studies show that this interface model is in the
same universality class as the TDE model and the qDDE model with
finite $q>2$. In section $8$ we propose some higher dimensional
generalizations of this interface model, which still conserve the
minimum height of the interface locally.
\section{Definition of the Interface model}
We consider the interface model on a one-dimensional lattice of size
$L$. At any given time, the interface is specified by the integer
height $h_i$ at each site $i$ of the lattice. The heights are
assumed to obey the restricted solid-on-solid (RSOS) condition
\begin{equation}
    h_{i+1}-h_i = \pm 1    
\end{equation}
for all $i$ and at all times. The interface evolves by the following
Markovian local dynamics: at every site the interface height can be
changed from $h_i$ to either $h_i+2$ or $h_i-2$ with some rates,
provided this will not violate the RSOS condition. {\it The rates
for the transition $h_i \rightarrow h_i \pm 2$ depends on the
next neighbouring heights $h_{i-2}$ and $h_{i+2}$}. 

\begin{figure}
\centerline{\psfig{figure=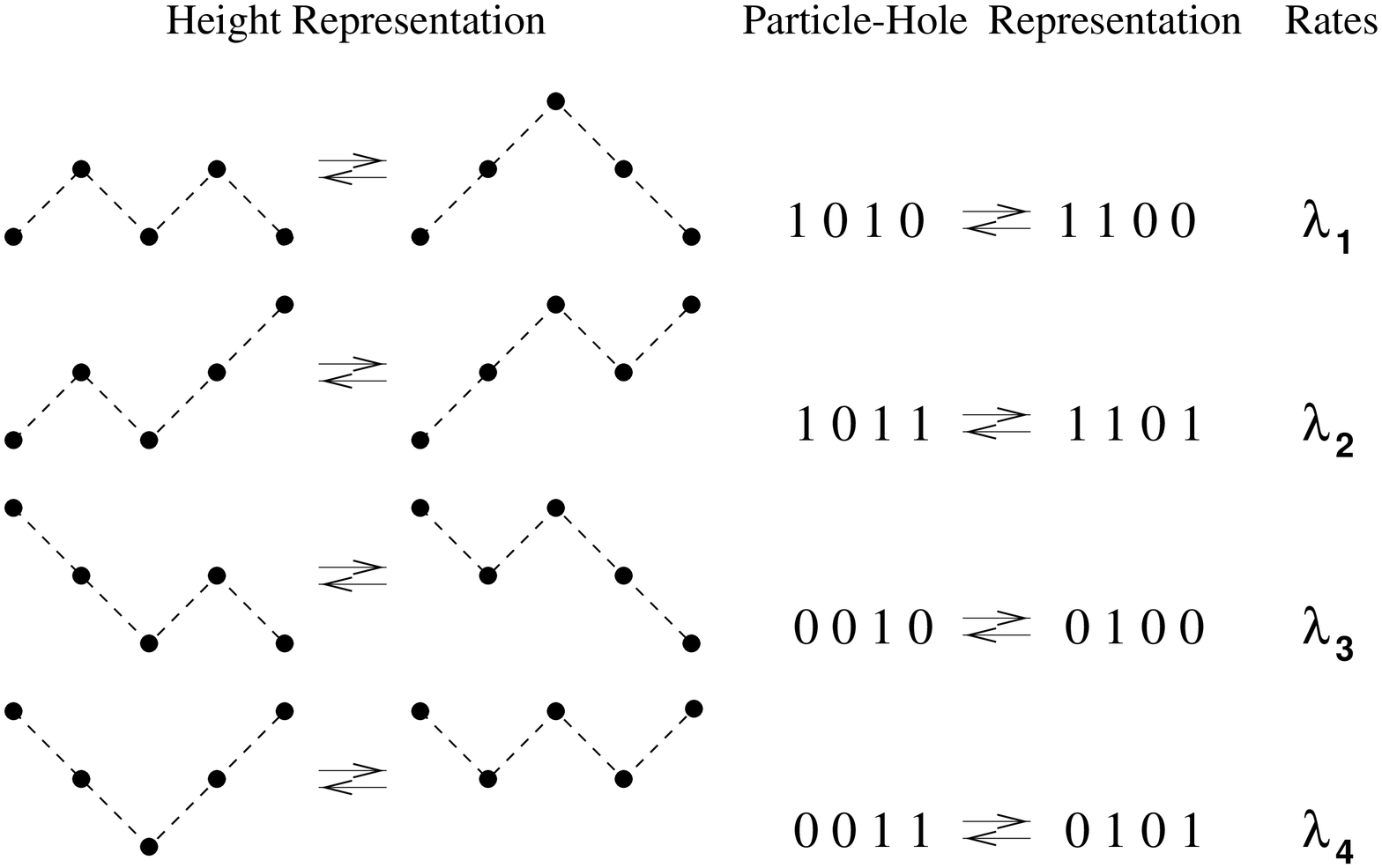,height=8cm}}
\caption {Transition rates of the  interface growth model.} \label{fig_3}
\end{figure}

Equivalently, we can specify the interface in terms of slope variables
$n_i= [h_{i+1}-h_i + 1]/2$, which takes only values $0$ and $1$.
Then change in the height at site $i$ corresponds to the exchange of
the variables $n$ at site $i$ and $i-1$. If we think of $n$'s as
occupation variables of a hard core lattice gas, this corresponds to
the well-known exclusion process, with hopping rates between sites $i$
and $i+1$ depends on the occupation at sites $i-1$ and $i+2$. In our
model we assume that the rate for the rightward and leftward hoppings
(or $h_i \rightarrow h_i +2$ and $h_i \rightarrow h_i-2$) are the
same. Then there are $4$ hopping rates, depending on the four possible
states of the sites $i-1$ and $i+2$. Let us call these rates
$\lambda_1,\lambda_2, \lambda_3$ and $\lambda_4$. These are shown in
figure \ref{fig_3}.

When the hopping rate between sites $i$ and $i+1$ is independent of
$n_{i-1}$ and $n_{i+2}$ (simple exclusion process), the model is
exactly solvable. The corresponding interface growth is described by
the Hammersley-Edwards-Wilkinson equation \cite{ham,edwk}. For this
model the dynamical exponent $z=2$. Further if the forward and
backward transition rates are not equal (asymmetric exclusion
process), the model would corresponds to a growing interface that is
described by the KPZ equation, for which $z=3/2$,
\cite{hhz,kpz}. The asymmetric version of our model is
interesting, but has not been studied in detail.

At time $t=0$, the interface height $h_i$ is given to be $0$ if
$i$ is odd, and $1$ if $i$ is even. At the boundaries, we can work
with fixed boundary conditions corresponding to choosing the height to
be $0$ at $i=0$ and $L+1$ at all times. Alternatively, we shall use
periodic boundary conditions so that $h_{L+1}=h_1$.

 This choice of rates satisfies the detailed balance condition, and in
the steady state, all allowed configurations occur with equal
weight. The case when forward and backward jumps occur with unequal
rates is of interest, but will not be discussed here.  If $ \lambda_i
{}= 1$ for all $i$, then this model corresponds to the Rouse model of
polymer dynamics \cite{rouse}.  If all the rates $\lambda_i,~i=1$ to
$4$ are nonzero, then it is easy to see that qualitative behavior of
the relaxation is not changed much. Let $\Delta
E(\lambda_1,\lambda_2,\lambda_3,\lambda_4)$ is the smallest nonzero
eigenvalue of the relaxation matrix. The relaxation matrix is
symmetrical, and can be thought of as the force matrix of a system of
mass points connected by springs with spring-constants $\lambda_i$. As
the eigenfrequencies are nondecreasing functions of the spring
constants, it follows that
\begin{equation}
     \frac{d}{d \lambda_i} \Delta E  \ge 0
\end{equation} 
It is then easy to deduce that
\begin{equation}
\lambda_{min}\Delta E(1,1,1,1)  
\le \Delta E(\lambda_1,\lambda_2,\lambda_3,\lambda_4)
\le \lambda_{max} \Delta E(1,1,1,1)
\label{XY}
\end{equation}
where $\lambda_{min}$ and $\lambda_{max}$ are the smallest and largest
values in a given $\{\lambda_i\}$.

Now, $\Delta E(1,1,1,1)$ is the lowest relaxation rate in the Rouse
model, and it is known that for a ring of size $L$, it decreases as
$L^{-2}$ for $L$ large. This implies that so long as
$\lambda_{min}$ and $\lambda_{max}$ are finite, $\Delta
E(\{\lambda_i\})$ decreases as $L^{-2}$ for large $L$, and the
dynamical exponent $z$ remains $2$.  However, if $\lambda_{min}$
becomes $0$, then there is a possibility that we get a different
universality class. {\it In this paper, we study the case when $\lambda_4
=0$, and all other $\lambda$'s are nonzero}. [ The case $\lambda_1 =
0$, all other $\lambda$'s nonzero is equivalent to this].  To keep the
interface model transition rules left-right symmetric, we assume
in addition that $\lambda_2=\lambda_3$.

An important consequence of $\lambda_4 = 0$ is that minimum of
$\{h_{i-2},h_{i-1},h_i,h_{i+1},h_{i+2}\}$ is conserved during a change
of heights at $i$. This implies, in particular, that this interface
dynamics conserves minimum of the full interface profile
$\{h_i\}$. {\it The constraint that minimum height is locally
conserved is a strong constraint that makes the dynamics slower than
the Rouse dynamics.} Note that the conclusion $z \ge 2$ for our model
follows immediately from the inequality \ref{XY}.
 
\begin{figure}
\begin{center}
\centerline{\psfig{figure=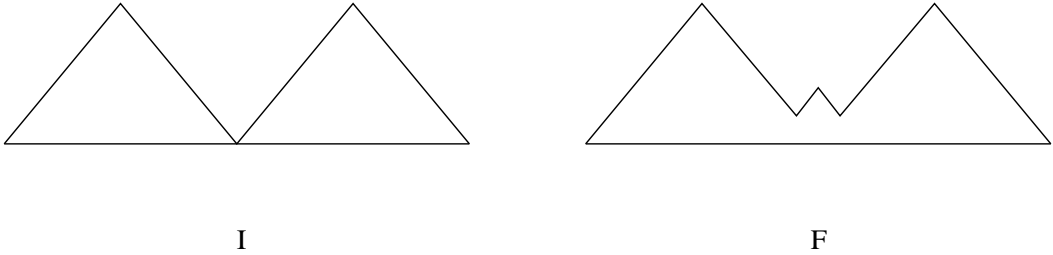,angle=270,width=14cm}}
\caption {An example of two configurations where it takes  order of
$L^2$ steps  to reach one from the other.}  \label{fig_4}
\end{center}
\end{figure}

As an illustration of this, consider the transition in figure
\ref{fig_4} from the initial configuration I to final configuration
F. Both these configurations are allowed, but to go from I to F, it
takes a long process of restructuring. The shortest route is to first
completely erase one of the two 'mountains', and then rebuild the
final structure F from that. It is easy to convince oneself that other
routes, essentially equivalent and requring as many steps, are
possible, but none requiring fewer can be found. For a pair of
mountains of size $L$ each, to go from I to F it takes order $L^2$
steps. In contrast in Rouse dynamics, F can be reached from I in one
step.

The constraint that minimum of height is locally conserved is
approximately realized in some physical situations, which give rise to
surface morphology sometimes known as `wedding cake' morphology with a
surface roughness exponent $\alpha$ close to 1 \cite{MBE}. These are
understood to be due to the existence of step-edge energy barriers of
the Ehrlich-Schwoebel type, which inhibit jumps of atoms from higher
steps to lower steps \cite{Krug}. Clearly, this makes growth at height
minima less likely. In the Monte Carlo simulations of Krug, the
surfaces generated have deep ridges which seem to survive for a long
time.  Our model is simpler than his, but generates shapes of surfaces
qualitatively similar. In particular, in both cases fluctuations in
height are of the same order as mean height.

It is quite striaghtforward, but instructive, to write the stochastic
matrix of the interface model as a quantum Hamiltonian $\hat{H}$. We
consider a chain of $L$ quantum mechanical spins $\{S_i\}, i = 1$ to
$L$. To each configuration $\{n_i\}$, we associate a spin
configuration $\{S^z_i\}$ by the rule $n_i=(1+S^z_i)/2$. Then it is
easily seen that \cite{alca} the quantum-mechanical Hamiltonian
corresponding to our model is
\begin{eqnarray}
\hat{H} &=& \sum_{i=1}^L f[\hat{S}_{i-1}^z,\hat{S}_{i+2}^z]~(\hat{\vec{S}}_i
\hat{\vec{S}}_{i+1} - 1),
\end{eqnarray}
where the function $f$ takes arbritary positive values except that
$f(-1,+1)=0$.  This is a four spin interaction Hamiltonian, which is
not yet tractable analytically. Note that the Hamiltonian is not
left-right symmetric though the original height model is. This is due
to the fact the transformation from $h_i$ to $n_i$ breaks the
reflection symmetry.  It is invariant under the simultaneous
interchange $x\longleftrightarrow -x$ and $n\longleftrightarrow 1-n$.

\section{Steady State Properties of the Interface}

Starting from an initial profile, the interface will grow until the
system reaches a steady state. The average interface height in the
steady state (the saturation height) can be computed exactly by
mapping the interface profiles to paths of a random walk that is not
allowed to cross the origin. Since the interface heights satisfy
$h_{i+1} = h_i \pm 1$, the set $\{h_i\}$ forms the path of a random
walk if we imagine $h$ and $i$ as space and time co-ordinates. The
constraint that the walk is not allowed to cross the origin comes from
the fact that the interface height at any point can not be negative.

Since the rules of the interface dynamics obey detailed balance, all
accesible interface configurations have equal weight in the steady
state. The average height of the interface then corresponds to the
average displacement of the random walk from the origin, this can be
computed exactly as follows: let us assume periodic boundary
conditions and $L=2n$, an even integer. We pick a point at random on
the ring, and call it the origin. The number of configurations where
height at the origin is $h$ is given by the number of paths from
$(i=0,h)$ to $(i=2n,h)$ which touches the line $h=0$ but donot fall
below. Let us denote this number by $\tilde N[(0,h) ; (2n,h)]$. If we
denote $N_{j}[(0,h);(2n,h)]$ the number of paths from $(0,h)$ to
$(2n,h)$ which donot fall below the line $h=j$,
\begin{equation}
\tilde N[(0,h) ; (2n,h)] = N_0[(0,h) ; (2n,h)]  
                                   - N_1[(0,h) ; (2n,h)] 
\end{equation}
{}From the well known reflection principle \cite{fell},
\begin{equation}
N_j[(0,h) ; (2n,h)] = M[(0,h) ; (2n,2j-h-2)]
\end{equation}
where $M[(0,h) ; (2n,h')]$ denotes the number of paths from
$(0,h)$ to $(2n,h')$ with no constraint on the path. Hence
\begin{equation}
\tilde N[(0,h) ; (2n,h)] = M[(0,h) ; (2n,-h)]
                                   - M[(0,h) ; (2n,-h-2)]
\end{equation}
Since 
\begin{eqnarray}
M[(0,h) ; (2n,h')] &=& ~{{2n}\choose {\frac{|h'-h|}{2}+n}}, \nonumber \\
\tilde N[(0,h) ; (2n,h)] &=& f(h) - f(h+1), 
\end{eqnarray}
where,
\begin{equation}
\hspace{1cm}f(h)= {{2n}\choose {n+h}} \nonumber
\end{equation}
Therefore the average height is given by
\begin{eqnarray}
\langle h \rangle &=& \frac{\sum_{h=0}^{n} h ~~[f(h) - f(h+1)]}{\sum_{h=0}^{n}
  [f(h) - f(h+1)]} \\ 
                  &=&\frac{1}{2} [ 2^{2n}/{{2n}\choose n} -1] 
\end{eqnarray}
which increases as $\sqrt{\frac{\pi}{8}}~L^{1/2}$ for large $L$.  For
an infinite system this means that starting from a profile which has
the minimum possible heights ($\{h_i\} = \{0,1,0,1,\ldots\}$), the
average interface height will always grow. However due to the
conservation of minimum height, it will always be tethered to $h=0$
at least at one point. This is an interesting example of a growing
interface with completely reversible rules.

\section{A Brief Review of the qDDE Model} 
\begin{figure}
\centerline{\psfig{figure=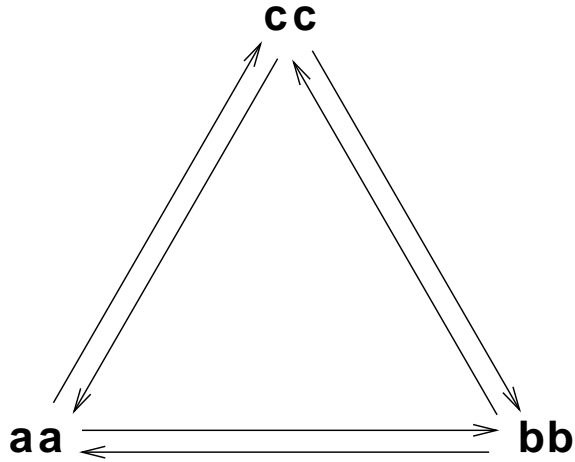,angle=270,height=6cm}}
\caption {Allowed transitions in the qDDE model for $q=3$}
\label{fig_1}
\end{figure}

The qDDE model is defined as follows: Consider a $d$-dimensional
lattice, where at each site there is a discrete variable which can
take $q$ distinct discrete values (colors). The system undergoes a
continuous time Markovian dynamics with the update rule that with rate
$1$ two neighboring spins having the same color, can simultaneously
change their color to any of the other ($q-1$) colors. For example
consider the case of $q=3$, and let $a,b$ and $c$ denote the $3$
colors. An $aa$ pair can become a $bb$ pair or a $cc$ pair with rate
$1$. In the same manner $bb$ and $cc$ pairs can change their
color. The dynamical rule for this case can be stated as in \hbox{Figure
\ref{fig_1}.}

The qDDE model has been studied in detail in $1+1$ dimensions
\cite{ddhmk2}. The main feature of the model is that its phase space
breaks up into an exponentially large number of dynamically
disconnected sectors. The number of sectors scales as $(q-1)^L$ with
system size. This strong nonergodic behavior is due to the presence of
a conserved quantity in the model, called {\it irreducible string}
(IS) which is defined as follows: A configuration of the qDDE model on
a linear chain of length $L$ can be represented by a string of $L$
characters where the $i$th character represents the color of the $i$th
spin. From this string delete all pairs of adjacent characters that
are the same.  Repeat this procedure on the resulting string until a
string with no immediate repetition of characters is obtained. This
string, whose length can not be reduced further by this reduction
algorithm is the IS corresponding to given configuration. It can be
shown that IS is conserved during the qDDE dynamics and each sector
has a distinct IS \cite{ddhmk2}. As a consequence, IS can be used to
uniquely specify a sector. The special sector for which the
irreducible string is a null string (the original string is completely
reducible) is called the {\it null sector} and is found to have a
slower relaxation to the steady state.

In any given sector, in the steady state all the configurations of
that sector occur with equal probability. This follows from detailed
balance. The number of configurations in each sector can be computed
exactly, and it typically grows as $\exp(L)$. For example, in the null
sector the number of configurations grows as $[4(q-1)]^{L/2}$.
 
In addition to IS, the qDDE model in one-dimension has another set of
conserved quantities which corresponds to a symmetry called {\it
recoloring symmetry} in the model. A qDDE configuration can be
represented by the configuration of a polymer chain on a
$q$-coordinated Bethe lattice.  To see this, consider a
$q$-coordinated Bethe lattice with its bonds colored with $q$
different colors such that all the bonds meeting at any given site
have different colors. We define the polymer chain corresponding to a
given qDDE configuration as the L step walk which starts from the
origin and follows the sequence of bonds such that their color is in
the same sequence as the colors in the qDDE configuration from site
$1$ to site $L$. Thus there is a one-to-one correspondence between the
L step polymer chain configurations on the Bethe lattice and the
configurations of the qDDE model. In this representation, the qDDE
dynamics corresponds to a kind of reptation motion of the polymer
chain on the Bethe lattice. A kink, consisting of two adjacent steps
of the polymer chain where an immediate retraversal occurs, can jump
to one of the neighbouring sites on the Bethe lattice. The rules of
the dynamics of the polymer chain is independent of the color of the
bonds and hence recoloring of the bonds is a symmetry of the
model. Note that this recoloring symmetry is not just the symmetry
under permutations of colors which corresponds to a global rotation in
the color space. It allows a local recoloring at each bond, and thus
is more like a gauge symmetry \footnote{Of course, we cannot change
the color only at a finite number of sites of the Bethe lattice
without violating the constraint that all bonds meeting at a site have
different colors}. Using the recoloring symmetry a large number of
eigenstates of the stochastic matrix of the model can be computed
exactly \cite{ddhmk2}. However these eigenstates are not the low lying
eigenstates, which determine the long time behavior and hence the
dynamical exponent $z$.

The decay of spin-spin correlation functions in the steady state is
found to be sector dependent for the qDDE model. The equal time two
point correlation function can be determined analytically, and in the
null sector it decays as $r^{-3/2}$. Hence we expect the time
dependent spin-spin auto-correlation function to decay as $t^{-3/{2
z}}$ asymptotically in this sector. Numerical diagonalization studies
of the stochastic matrix shows that the dynamical exponent $z\approx
2.5$ in the null sector.  In other sectors the decay of the
autocorrelation function can be determined using the relationship to
hard core random walk with conserved spin (HCRWCS) model introduced in
\cite{ddmb3}.  HCRWCS model corresponds to a simple exclusion process
with the particles carrying a spin label on each of them. Hence for
these non-null sectors the dynamical exponent is $2$.

\section{Dynamics in the Large q Limit}
\begin{figure}
\centerline{\psfig{figure=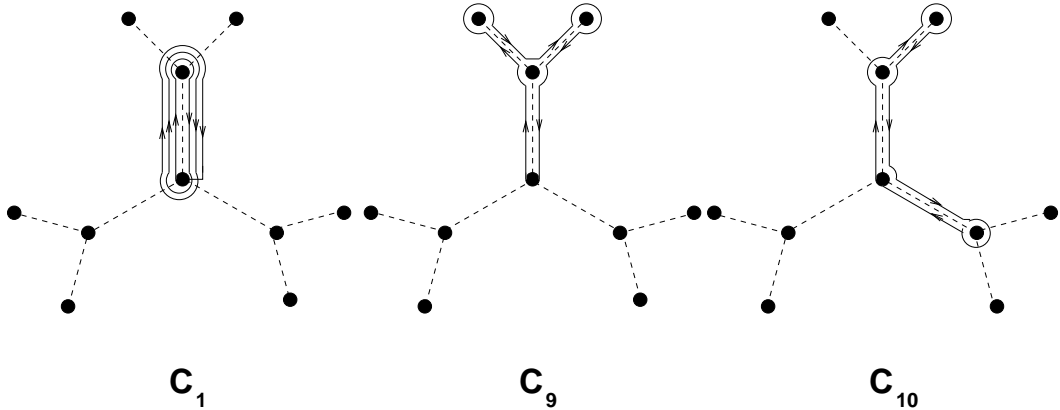,width=14cm}}
\caption{Polymer chain configurations on the Bethe lattice 
corresponding to equivalence classes $\mathbb{C}_1$, $\mathbb{C}_9$
and $\mathbb{C}_{10}$ of the qDDE model on a chain of length
$L=6$. For every site only $3$ bonds are shown.}\label{fig_2}
\end{figure}

Consider the qDDE model on a chain of length L with open boundary
conditions. Here we restrict to only the null sector but our following
analysis is equally applicable to other sectors. In terms of polymer
chain configurations, we consider the case where both the end points
of the chain are fixed to be at the origin.  We divide the set of
configurations in this sector into equivalence classes such that all
configurations in a given equivalence class are related to each other
by recoloring symmetry. As an example, in table \ref{table_1} all the
equivalence classes are shown for $L=6$ along with the number of
configurations in each class.  Each class has a unique topology of
polymer chain configuration on the Bethe lattice. As an example the
polymer chain configurations of equivalence classes $\mathbb{C}_1$,
$\mathbb{C}_9$ and $\mathbb{C}_{10}$ are shown in figure \ref{fig_2}.

\begin{table}{}
\begin{center}
\begin{tabular}{|l|l|l|} 
\hline 
Equivalence & Representative & Number of \\
Class      & Element        & Configurations \\
\hline 
\hspace{0.5cm}$\mathbb{C}_1$ & \hspace{0.5cm}aaaaaa & q\\
\hspace{0.5cm}$\mathbb{C}_2$ & \hspace{0.5cm}aaaabb & q (q-1)\\
\hspace{0.5cm}$\mathbb{C}_3$ & \hspace{0.5cm}aaabba & q (q-1)\\
\hspace{0.5cm}$\mathbb{C}_4$ & \hspace{0.5cm}aabbaa & q (q-1)\\
\hspace{0.5cm}$\mathbb{C}_5$ & \hspace{0.5cm}abbaaa & q (q-1)\\
\hspace{0.5cm}$\mathbb{C}_6$ & \hspace{0.5cm}bbaaaa & q (q-1)\\  
\hspace{0.5cm}$\mathbb{C}_7$ & \hspace{0.5cm}abbbba & q (q-1)\\
\hspace{0.5cm}$\mathbb{C}_8$ & \hspace{0.5cm}aabbcc & q (q-1) (q-2) \\
\hspace{0.5cm}$\mathbb{C}_9$ & \hspace{0.5cm}abbcca & q (q-1) (q-2)\\
\hspace{0.5cm}$\mathbb{C}_{10}$& \hspace{0.5cm}abbacc & q (q-1) (q-1)\\
\hspace{0.5cm}$\mathbb{C}_{11}$& \hspace{0.5cm}bbacca & q (q-1) (q-1)\\
\hspace{0.5cm}$\mathbb{C}_{12}$& \hspace{0.5cm}abccba & q (q-1) (q-1)\\
\hline
\end{tabular} 
\end{center}
\caption{Equivalence classes  under recoloring symmetry of the qDDE
model in the null sector for $L=6$. Here $a,b,c,\ldots$ represent
different colors. Colors which are adjacent but represented by
different symbols are assumed to be distinct. For example in
$\mathbb{C}_{10}$, $a \ne b$ and $a \ne c$, but $b$ and $c$ need not
be distinct. \label{table_1}} 
\end{table}

Note that for $\mathbb{C}_8-\mathbb{C}_{12}$, each bond of the Bethe
lattice is traversed either twice or not at all by the polymer chain
and these classes have more number of configurations than the other
classes. In general if the polymer chain occupy $n$ distinct bonds of
the Bethe lattice, then the number of configurations in the
corresponding equivalence class is approximately $q^n$ for large
$q$. As in the steady state all configurations appear with equal
probability, the weight of an equivalence class in the steady state is
directly proportional to the number of configurations in it. This
implies that in the $q \rightarrow \infty$ limit, only those
equivalence classes where the polymer chain traverse each bond of the
Bethe lattice exactly twice or not at all, will have a nonzero weight
in the steady state.  For example in the $L=6$ case, only the
equivalence classes $\mathbb{C}_8-\mathbb{C}_{12}$ will have a nonzero
weight in the steady state in the $q \rightarrow \infty$ limit. If the
state of the chain is examined at some instant in the steady state,
with probability $1$ it will belong to one of the equivalence classes
$\mathbb{C}_8-\mathbb{C}_{12}$. And the weight of all these
equivalence classes are the same in this limit, which we can take as
$1$. Let $\mathbb{L}$ denotes any equivalence class that has weight
$1$ in the steady state and $\mathbb{S}$ denotes any equivalence class
whose weight tends to $0$ in the steady state, in the limit of large
$q$.

In the steady state, a finite fraction of spins can flip at any
instant. For the spin at site $i$ to flip it is necessary that at
least one of its neighbors (at site $i-1$ or $i+1$) should have the
same color. The probability for this is given by $P = 2 q
N_{L-2}/N_L$, where $N_L$ is the number of configurations in the null
sector on a lattice of length $L$. As $N_L \sim [4 (q-1)]^{L/2}$ for
large $L$, $P=1/2$ in the limit $q \rightarrow \infty$.

We have already seen that, in the large $q$ limit, with probability
$1$ the equivalence class of a configuration in the steady state is an
$\mathbb{L}$. The average time the system spends in a particular
configuration is very small, of order $1/qL$. This is because in a
typical configuration there are order of $L$ flippable pairs, and each
such pair can go to any of the approximately $q$ other states with
rate $1$. However most of these transitions are within the same
equivalence class. For example, consider a local configuration of a
flippable pair $|\ldots abbc \ldots\rangle$. If the pair $bb$ changes
its color to $d$ which is different from both $a$ and $c$, then the
resulting configuration is equivalent to the old by recoloring
symmetry. On the otherhand the equivalence classes $\mathbb{S}$ are
very short lived. A local configuration of the type $|\ldots aaac
\ldots\rangle$ within a time of order $1/q$ will revert back to one of
the type $|\ldots abbc \ldots\rangle$ or $|\ldots bbac \ldots\rangle$.

Thus, for large $q$, most of the dynamics of qDDE model involves
transitions within an equivalence class of type $\mathbb{L}$.  The
only transitions allowed from an $\mathbb{L}$ equivalence class to a
different equivalence class is to one of type $\mathbb{S}$.  These
occur with rate of order $1$. Let us say that there is an allowed
transition from equivalence class $\mathbb{C}$ to one $\mathbb{C}'$,
where $\mathbb{C}$ is of type $\mathbb{L}$ and $\mathbb{C}'$ is of
type $\mathbb{S}$. $\mathbb{C}'$ being short lived, in a time of order
$1/q$ it makes a transition to a $\mathbb{C}''$ of type
$\mathbb{L}$. Let us say it goes to longlived classes
$\mathbb{C}''_1,\mathbb{C}''_2,\ldots$ with probabilities
$p_1\Gamma,p_2\Gamma,\ldots$. In this way, in the limit of large $q$,
for times $>>1/q$, we have a coarsegrained description of the
stochastic evolution of the qDDE system as transitions between
different longlived equivalent classes, and this effective dynamics is
{\it Markovian} with specified rates.
\section{Equivalence to the Interface Model} 
We now show that the effective dynamics of the qDDE model in the large
$q$ limit is equivalent to the dynamics of the interface model defined
in section 2.  Let $\mathsf{S}_i(C)$ denotes the substring
corresponding to sites from $1$ up to and including $i$ of a
configuration $C$ of the qDDE model. And let $h_i(C)$ be the length of
the IS corresponding to this substring $\mathsf{S}_i(C)$.  It is easy
to see that $h_i$'s are non-negative integers and $h_{i+1} - h_i = \pm
1$. The set $\{h_i(C)\}$ will be the same for all configurations $C$
in the same equivalence class $\mathbb{C}$. Hence the set $\{h_i(C)\}$
is a function of only the equivalence class $\mathbb{C}$ and we may
write it as $\{h_i(\mathbb{C})\}$. Moreover the correspondence between
equivalence classes $\mathbb{L}$ and $\{h_i\}$ is one to one. Hence
every equivalence class $\mathbb{L}$ can be uniquely represented by a
set $\{h_i\}$.  We identify ${h_i}'s$ with the height variables of the
interface model defined in Section 2. Then, the transitions between
these equivalence classes in the qDDE model give rise to a Markovian
time-evolution of the interface model.

We now proceed to derive the transition rates for this interface
dynamics and show that they correspond to special values of 
$\lambda_1,\lambda_2$ and $\lambda_3$. 

As $h_i$ represents the length of the IS of the substring up to site
$i$, $h_i$ changes only when the spins at sites $i$ and $i+1$ in the
qDDE model are flipped. These can happen only if they have the same
color, in which case $h_{i-1}=h_{i+1}$. Again, as $|h_{i+1} - h_{i}| =
1$, the only allowed transitions are of the type
\[
\{\ldots h, h-1, h, \ldots \} ~~\longleftrightarrow~~ \{ \ldots h,
h+1, h, \ldots \}
\]
The rates for this transition will depend on the second neighboring
heights $h_{i-2}$ and $h_{i+2}$. To find the transition rates, let
$\mathsf{S}$ be the local color configuration at the sites $i-2$ to
$i+2$. We write $\mathsf{S}$ as $s_1s_2s_3s_4s_5$, where
$s_1,s_2,s_3,s_4$ and $s_5$ represents the color at sites
$i-2,i-1,i,i+1$ and $i+2$ respectively. Each $s_i$ can be any of the
colors $a,b,c,d \ldots$ and their different combinations will
correspond to different local height configurations $\{h_i\}=\{\ldots,
h_{i-2},h_{i-1},h_i,h_{i+1},h_{i+2},\ldots\}$. We discuss them one by
one.

\noindent {\bf Case I.{ $ \{h_i\} = \{\ldots,h,h+1,h,h+1,h,\ldots \}$}}

In this case $\mathsf{S}$ is given by $abbcc$. Consider the transition
$cc \rightarrow bb$. As a result $\mathsf{S}$ becomes $abbbb$. As
explained earlier, this state is very short lived.  There are $3$ pairs
of $b$ that can flip. It is easy to see that flipping the first or the
third $bb$ pair does not change $h_i$. Flipping the middle $bb$ pair
increases $h_i$ to $h_i + 2$.  Since all the $3$ $bb$ pairs have equal
chance of flipping, the effective rate for the middle pair to flip is
$1/3$.  There is another possible way for $h_i \rightarrow h_i +
2$. Starting from $abbcc$, first $bb$ flips to $cc$ and then the middle
$cc$ pair flips. Thus the net effective rate for $h_i \rightarrow h_i
+ 2$ is $2/3$ in this case.

\noindent {\bf Case II. {$ \{h_i\} =
\{\ldots,h+2,h+1,h,h+1,h,\ldots \}$}}
   
In this case $\mathsf{S}$ is of the type $abcdd$ such that to the
left of $a$ there is one of each $a, b$ and $c$ with fully reducible
substrings in between them. Consider the transition $dd \rightarrow
cc$. The new $\mathsf{S}$ is $abccc$. The flipping of the first $cc$
pair will increase $h_i$ by $2$. Whereas flipping of the second $cc$
pair will not change $h_i$. Since both these pairs have equal chance
of flipping, the rate for $h_i \rightarrow h_i + 2$ is $1/2$ in this
case.

\noindent {\bf Case III. {$ \{h_i\} =
\{\ldots,h,h+1,h,h+1,h+2,\ldots \}$}}

For these heights $\mathsf{S}$ is of the type $abbcd$.
The only sequence of transitions which changes $h_i$ is first $bb
\rightarrow cc$, followed by the flipping of second $cc$ pair.
As in the previous case the rate for this transition, which 
increases $h_i$ by $2$, is $1/2$ for this case.

\noindent {\bf Case IV. {$ \{h_i\} =
\{\ldots,h+2,h+1,h,h+1,h+2,\ldots \}$}}

For this case $\mathsf{S}$ is of the type $abcde$ with no colors
equal. Hence the transition rate out of this local configuration is
zero.

The remaining four cases corresponds to reverse transitions ($h_i
\rightarrow h_i - 2$) of the above four cases. For example, if the
initial heights are $\{h_i\} = \{\ldots,h-2,h-1,h,h-1,h-2,\ldots \}$,
$h_i$ can only decrease by $2$ and it corresponds to the reverse
transition of case I. Arguing as before, it is easy to determine the
rates of these reverse transitions, and they are found to be the same
as that of the corresponding $h_i \rightarrow h_i+2$ transition.
Thus the qDDE model in the large $q$ limit corresponds to the
interface model of section $2$ with $\lambda_1 = 2/3,
\lambda_2=\lambda_3=1/2$

\section{Monte Carlo and Numerical Diagonalization 
Studies of the Interface Model} We have studied the interface model by
using both Monte Carlo simulations and exact numerical diagonalization
of the stochastic matrix. Monte Carlo simulations shows that the
average height of the interface shows a scaling form
\begin{equation}
\langle h(t,L) \rangle \sim L^{\alpha} f(t/L^z)
\label{eq4_1}
\end{equation}
where $L$ is the length of the lattice. The scaling function $f(x)
\rightarrow x^{\alpha/z}$ as $x \rightarrow 0$ and it become
a constant in the limit $x \rightarrow \infty$. In 
section 3 we have shown that $\alpha = 1/2$. 

\begin{figure}
\centerline{\psfig{figure=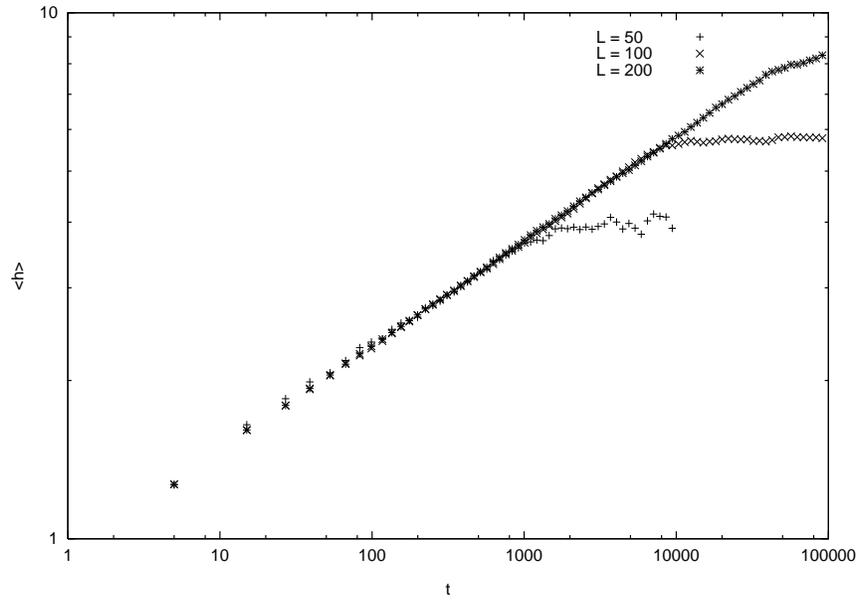,height=8cm}}
\caption{Average height $\langle h \rangle$ of the interface as
a function of time  for various lattice sizes.}
\label{fig_5}
\end{figure}

\begin{figure}
\centerline{\psfig{figure=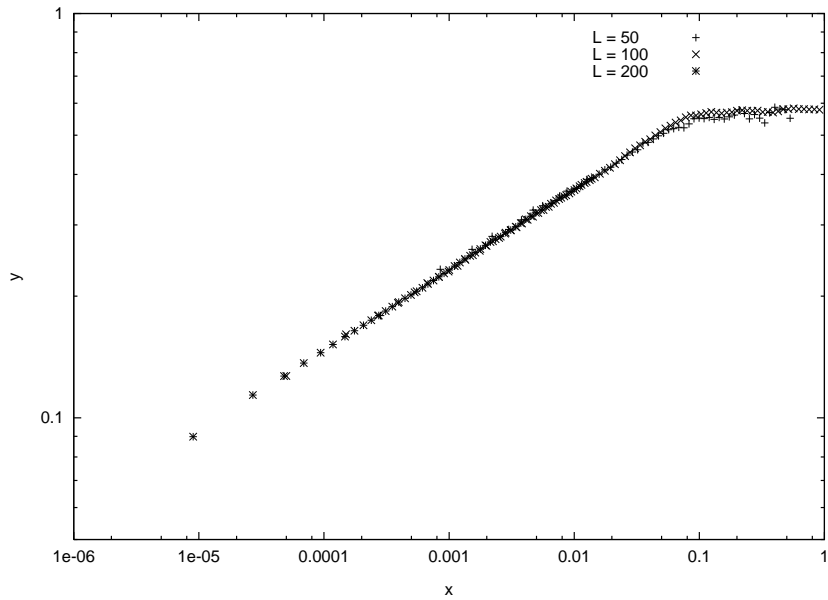,height=8cm}}
\caption{Collapse of the various curves in figure \ref{fig_5}. 
$y = \langle h \rangle L^{-1/2}$ is plotted  as a function of the
scaling variable $x = t/L^{5/2}$ for various lattice sizes.}
\label{fig_6}
\end{figure}

We have done Monte Carlo simulations for various lattice sizes.
These are shown in \hbox{Figure \ref{fig_5}}. These can be collapsed
into a  singe curve using the scaling form (\ref{eq4_1}) with
$\alpha=1/2$ and the dynamical exponent $z=2.5$. This is shown
in figure \ref{fig_6}.

We have also determined the dynamical exponent $z$ by numerical
diagonalization of the stochastic matrix for finite rings and
extrapolating the results to infinite $L$.

Note that we diagonalized the stochastic matrix of the interface model
for a ring of $L$ sites using periodic boundary conditions. This,
however, does not correspond to $q\rightarrow \infty$ limit of the
qDDE model on a ring of size $L$. This is because the one-to-one
correspondence between the equivalence classes of the qDDE model and
the interface configurations was proved only with fixed bounadry
conditions. In fact, many distinct configurations of the height model
on a ring can correspond to same equivalence class under recoloring of
the qDDE model on a ring. For example, for $L=6$, there are only two
distinct equivalence classes of qDDE model ( corresponding to
$\sqsubset$ or Y shaped ring polymers), but the height model with
periodic boundary conditions has $10$ configurations. The difference
between these models should however not be important for large $L$.

\begin{figure}
\centerline{\psfig{figure=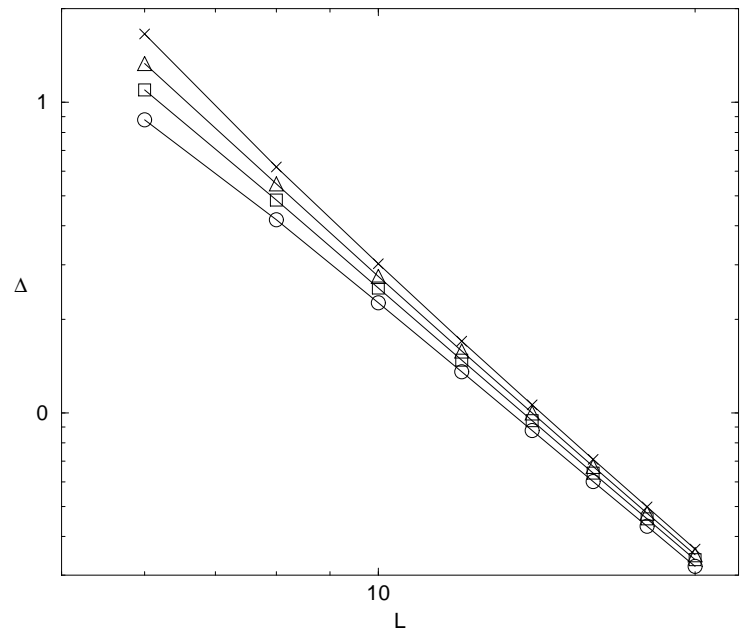,height=10cm,angle=270}}
\caption{Plot of $\Delta$ vs $L$ on a Log-Log scale for
$\lambda_2=\lambda_3=1$ and different values of 
$~\lambda_1 = 0.75 ~(\circ),~1.0~(\Box),~1.3333~(\triangle),~2~(\times)$.
Numerically determined values of the slope of these lines are
$2.82,2.90,2.96$ and $3.05$}
\label{fig_7}
\end{figure}

\begin{figure}
\centerline{\psfig{figure=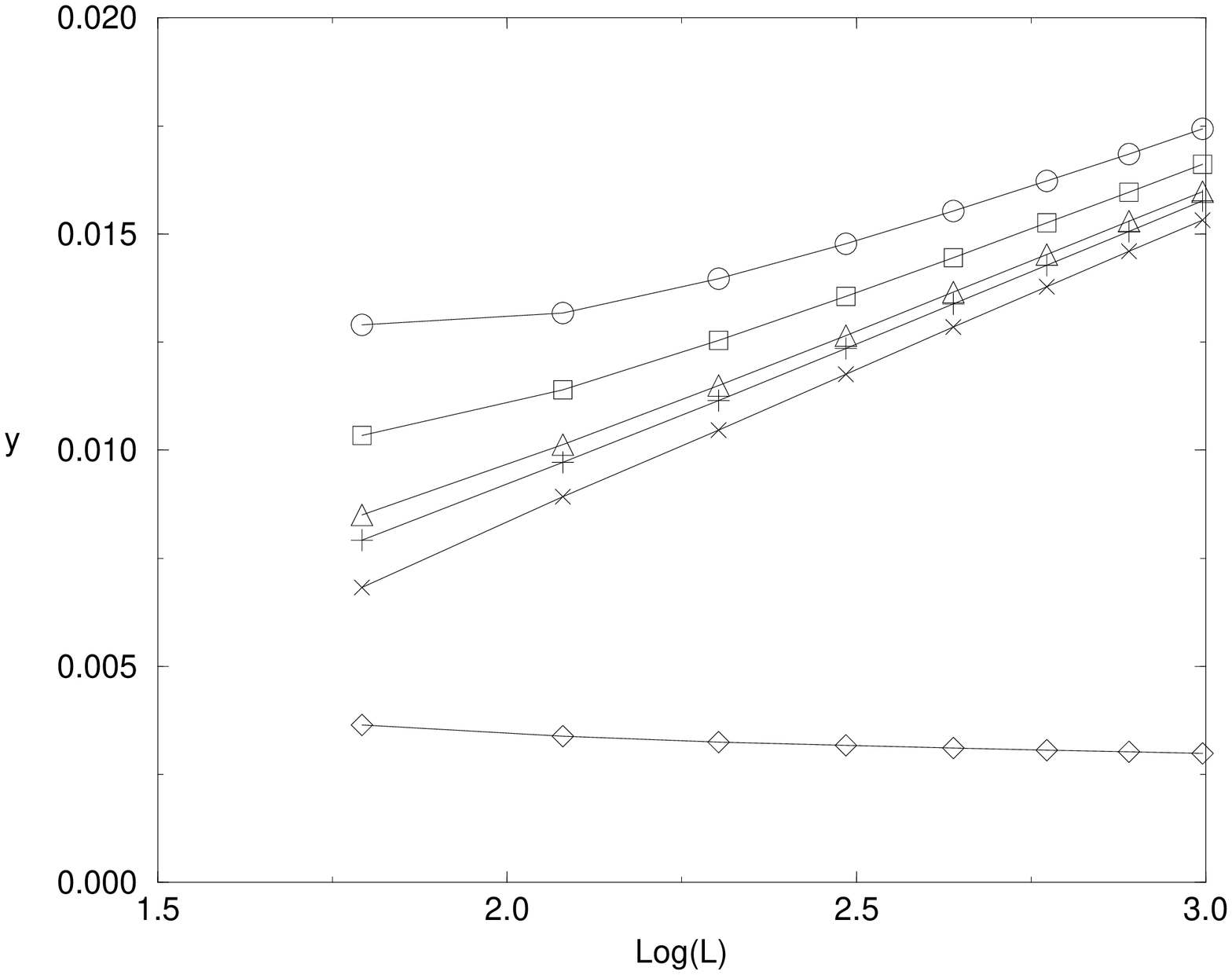,height=10cm,angle=270}}
\caption{Plot of $y=1/(\Delta~L^{2.5})$ vs $\log(L)$ for 
A) $\lambda_2=\lambda_3=1$ and different values of 
$~\lambda_1 = 0.75 ~(\circ),~1.0~(\Box),~1.3333~(\triangle),~1.5~(+),
~2~(\times)$, B) for the $3$-color qDDE model ($\diamond$)}
\label{fig_8}
\end{figure}

Our procedure of numerical diagonalization is very similar to the one
we have used in an earlier study of the TDE model \cite{ddhmk1}. Here
we only briefly out line our procedure. First we reduce the size of
the matrix to be diagonalized using symmetries like translation and
reflection. Then we find the second largest eigenvalue of the
stochastic matrix by numerical diagonalization. Since the largest
eigenvalue is zero, this will give the gap $\Delta $ in the eigenvalue
spectrum.  Assuming that the gap scales with system size as $\Delta
\sim L^{-z}$, we define an effective dynamical exponent,
\begin{equation}
z_L = \frac{\log[\Delta_{L-2}/\Delta_{L}]}{\log[L/(L-2)]}
\end{equation}
where $\Delta_L$ is the gap for a ring of size $L$. The true dynamical
exponent $z$ is then obtained by extrapolating $z_L$ to $L=\infty$.
We were able to go up to L=20. The size of the matrix to be
diagonalized is much smaller than that for the $q=3$ case of the qDDE
model \cite{ddhmk2}, a simplification achieved in the $q\rightarrow
\infty$ limit. We diagonalized the stochastic matrix for various
values of $\lambda_1$ keeping $\lambda_2=\lambda_3=1$.  The exponent
obtained by extrapolating $z_L$ to $L=\infty$ is about $2.8$, and it
depends on $\lambda_1$.  In figure (\ref{fig_7}), we have plotted the
values of $\Delta (L)$ versus $L$ on a log-log plot for various values
of $\lambda_1$. The data donot falls into parallel straight lines.
Assuming that this is due to corrections to the
scaling of the gap, we tried scaling of the form
\begin{equation}
\Delta(L)= A L^{-z_1} + B L^{-z_2}
\end{equation}
The best fit were obtained for values of $z_1$ and $z_2$ both very close to
$2.5$. This suggested the extrapolation form
\begin{equation}
\Delta \sim L^{-z}/\log(L/a)
\label{eq_sc}
\end{equation}
In figure (\ref{fig_8}) we have plotted $1/(\Delta L^{2.5})$ vs
$\log(L)$ for various values of $\lambda_1$, keeping
$\lambda_2=\lambda_3=1$. Note that the plot involves no fitting
parameters, except for the exponent $2.5$. The constant $a$ in
equation (\ref{eq_sc}) just gives an overall shift to the plot and
hence can be taken to be $1$. We find that all points falls into
straight lines, which gives us some confidence that equation
(\ref{eq_sc}) is likely to be the correct scaling form.  Note that the
$\log(L)$ correction should modify the scaling of the average
interface height (equation (\ref{eq4_1})) to $\langle h(t,L) \rangle
\sim L^{\alpha} f[t/(L^z \log(L/a))]$. We have tried this scaling form
for $z=2.5$ and $a=1$, and it gave reasonably good collapse, though
best collapse seems to be obtained for a slightly lower value of z about
$2.4$.

To see if the logarithmic correction term is responsible for the poor
convergence of observed effective dynamical exponent in the earlier
exact diagonalization study of the TDE and $3$-color qDDE models
\cite{ddhmk1,ddhmk2}, we have also plotted earlier data for the
$3$-color qDDE model in figure (\ref{fig_8}). The graph is again
fairly linear with very small {\it negative} slope. As a negative
slope is not possible asymptotically, we conclude that the 3-DDE model
shows no evidence of a logarithmic correction term.  Graph for the gap
in the TDE model is also quite similar (not shown here). This is
consistent with the expectation that these models are in the same
universality class, upto logarithmic corrections.

\section{Higher Dimensional Generalizations}
It is easy to construct higher dimensional interface growth models
having the property that minimum of heights is conserved
locally. These are however may not realizable as the $q \rightarrow \infty$
limit of higher dimensional qDDE models. 

Consider an interface model on a square lattice where the
heights $h(i,j)$ are integers and nearest neighbor slopes takes values
$\pm 1$. The growth rule is that $h(i,j) \rightarrow h(i,j) + 2$, if
all neighbors have height $h(i,j) + 1$ and at least one of the $4$
second neighbors ($(i\pm2,j),(i,j\pm2)$) has height $h(i,j)$. Reverse
transition takes place with the same rate. In this model also
Min($\{h(i,j)\}$) is conserved. Mountain craters analogue of that in
Figure \ref{fig_4} are easy to construct such that it takes a long
process of restructuring to go from one to another. 

Another simple variant of this model which avoids the odd-even
sublattices is the following: We consider an RSOS model with an
integer height coordinate $h(i,j)$ at the site $(i,j)$ of a square
lattice. The difference in heights at two adjacent sites is
constrained to be one of three values: $-1, 0$ or $1$.  Thus we allow
neighboring sites to have the same height. In the initial
configuration, $h(i,j)=0$ for all sites $(i,j)$.  The transition rule is
that $h(i,j) \rightarrow h(i,j) + 1$ with rate $1$, if this would not
violate the RSOS constraint, and if at least one of the neighbors have
the same height as $(i,j)$. The reverse transition also occurs at the
same rate. Clearly this also gives rise to a fluctuating surface whose
minimum height value does not change in time.  A detailed
investigation has not yet been undertaken.  
\vspace{0.5cm}

\noindent Acknowledgment: We would like to thank M Barma,
G I Menon and J. Krug for some useful discussions.
\vspace{0.5cm}

\noindent \hbox{$\dagger$Present Address:Department of Physics of Complex 
Systems,}\linebreak \hbox{Weizmann Institute of Science}, \linebreak 
\hbox{Rehovot 76100, Israel.}\linebreak 
\hbox{$\dagger$E-mail:fehari@wicc.weizmann.ac.il}\linebreak  
\hbox{$\ddagger$E-mail:ddhar@theory.tifr.res.in}

\end{document}